\documentclass{PoS}

\usepackage{tikz}
\usepackage{amsmath,amsthm}
\usepackage{amssymb}
\usepackage[numbers, sort&compress]{natbib}
\def\d{{\rm d}}

\title{Differential equations for loop integrals without squared propagators}

\ShortTitle{Differential equations for loop integrals without squared propagators}

\author{Jorrit Bosma\\
        ETH Z{\"u}rich, Wolfang-Pauli-Strasse 27, 8093 Z{\"u}rich, Switzerland\\
        E-mail: \email{jbosma@itp.phys.ethz.ch}}

\author{\speaker{Kasper J. Larsen}\\
        School of Physics and Astronomy, University of Southampton, Highfield, Southampton, \\
        SO17 1BJ, United Kingdom\\
        E-mail: \email{Kasper.Larsen@soton.ac.uk}}

\author{Yang Zhang\\
        ETH Z{\"u}rich, Wolfang-Pauli-Strasse 27, 8093 Z{\"u}rich, Switzerland\\
        PRISMA Cluster of Excellence, Johannes Gutenberg University, 55128 Mainz, Germany\\
        E-mail: \email{zhang@uni-mainz.de}}

\abstract{We provide a sufficient condition for avoiding
squared propagators in the intermediate stages
of setting up differential equations for loop integrals.
This condition is satisfied in a large class of two- and three-loop diagrams.
For these diagrams, the differential equations can
thus be computed using ``unitarity-compatible''
integration-by-parts reductions, which simplify
the reduction problem by avoiding integrals with higher-power propagators.}

\FullConference{Loops and Legs in Quantum Field Theory (LL2018)\\
		29 April 2018 - 04 May 2018\\
		St. Goar, Germany}

\begin{document}

\section{Introduction}\label{sec:introduction}

The physics program of the Large Hadron Collider (LHC)
demands precision calculations of cross sections
of Standard Model processes to attain a quantitative
understanding of the background and in turn facilitate
the extraction of new physics signals. The required accuracy
is typically next-to-next-to leading order (NNLO)
in fixed-order perturbation theory, in order to match
the parton distribution uncertainties and the experimental precision.
Out of the contributions to NNLO cross sections, the double-virtual one,
i.e.~the two-loop scattering amplitude, is the main bottleneck.

Calculation of multi-loop amplitudes proceeds in two stages. In the first step,
the amplitude is rewritten as a linear combination of a
basis of integrals through the use of discrete symmetries
and integration-by-parts (IBP) reductions. The latter are
linear relations among loop integrals which arise from
the vanishing integration of total derivatives in dimensional regularization,
\begin{equation}
\int \prod_{i=1}^L \frac{\d^D \ell_i}{\mathrm{i} \pi^{D/2}}
\sum_{j=1}^L \frac{\partial}{\partial \ell_j^\mu}
\frac{v_j^\mu \hspace{0.5mm} P}{D_1^{\alpha_1} \cdots D_m^{\alpha_m}}
\hspace{1mm}=\hspace{1mm} 0 \,, \label{eq:IBP_schematic}
\end{equation}
where $P$ and the vectors $v_j^\mu$ are polynomial in the internal and
external momenta, the $D_k$ denote inverse propagators, and the
$\alpha_i$ are integers. Upon applying Gaussian elimination~\cite{Laporta:2000dc,Laporta:2001dd}
to a suitably large system of IBP identities (\ref{eq:IBP_schematic}),
one obtains the IBP reductions~%
\cite{Anastasiou:2004vj,Smirnov:2008iw,Smirnov:2014hma,Studerus:2009ye,%
vonManteuffel:2012np,Lee:2012cn,Maierhoefer:2017hyi,Gluza:2010ws,Ita:2015tya,%
Larsen:2015ped,vonManteuffel:2014ixa,Bern:2017gdk,Boehm:2018fpv,Chawdhry:2018awn},
which express the majority of the contributing
loop integrals as linear combinations of a small basis of integrals.

In the second step, one sets up differential equations
for the basis integrals~%
\cite{Kotikov:1990kg,Kotikov:1991pm,Bern:1993kr,Remiddi:1997ny,Gehrmann:1999as,Henn:2013pwa,%
Papadopoulos:2014lla,Ablinger:2015tua,Frellesvig:2017aai,Zeng:2017ipr}. Letting $x_m$
denote an external kinematical invariant, $\epsilon = \frac{4-D}{2}$ the
dimensional regulator, and $\boldsymbol{\mathcal{I}}(\boldsymbol{x},\epsilon)
= (\mathcal{I}_1(\boldsymbol{x},\epsilon), \ldots, \mathcal{I}_M(\boldsymbol{x},\epsilon))$
the basis of integrals, we have the following first-order linear system,
\begin{equation}
\frac{\partial}{\partial x_m} \boldsymbol{\mathcal{I}}(\boldsymbol{x},\epsilon)
= A_m (\boldsymbol{x}, \epsilon) \boldsymbol{\mathcal{I}}(\boldsymbol{x},\epsilon) \,,
\label{eq:diff_eqs_schematic}
\end{equation}
where in practice one uses the IBP reductions to decompose
the derivatives $\frac{\partial \mathcal{I}_j}{\partial x_m}$
in the basis. With appropriate boundary conditions,
eq.~(\ref{eq:diff_eqs_schematic}) can
be solved to produce expressions for the basis integrals.
This method has proven to be a powerful tool
for computing multi-loop integrals.

The aim of these proceedings, based on ref.~\cite{Bosma:2017hrk},
is to investigate to what extent the IBP reduction formalisms
of refs.~\cite{Gluza:2010ws,Ita:2015tya,Larsen:2015ped,Boehm:2018fpv}
are compatible with differential equations of the type in
eq.~(\ref{eq:diff_eqs_schematic}). The main idea of
these IBP reduction formalisms
is to choose the $v_j^\mu (\ell_i)$
in eq.~(\ref{eq:IBP_schematic}) such that the resulting IBP identities
do not involve squared propagators. The resulting IBP identities
thus involve a more limited set of integrals and therefore
produce significantly smaller linear systems to be solved.
The question we wish to address is therefore whether
it is possible to set up differential equations
of the form (\ref{eq:diff_eqs_schematic}) without generating
integrals with squared propagators in intermediate stages.

\section{Differential equations in Baikov representation}\label{sec:Baikov_representation}

We begin by fixing our notation and conventions.
We consider a Feynman integral with $L$ loops, $k$ propagators and
$m-k$ irreducible scalar products (i.e., polynomials in the
loop momenta and external momenta which cannot be expressed
as a linear combination of the inverse propagators).
We apply dimensional regularization and normalize the
integral as follows,
\begin{equation}
I(N; \alpha_1, \ldots, \alpha_m; D) \equiv \int \prod_{j=1}^L \frac{\d^D \ell_j}{\mathrm{i} \pi^{D/2}}
\frac{N}{D_1^{\alpha_1} \cdots D_m^{\alpha_m}} \,. \label{eq:def_generic_Feynman_integral}
\end{equation}
Here $N$ denotes a polynomial in the linearly independent
external momenta $p_1, \ldots, p_E$ and the loop momenta $\ell_1, \ldots, \ell_L$,
and $m=LE + L(L+1)/2$. The propagators are labeled such that,
\begin{align}
\alpha_i & \geq 1 \hspace{5mm} \mathrm{for} \hspace{5mm} i=1,\ldots, k \nonumber \\
\alpha_i & \leq 0 \hspace{5mm} \mathrm{for} \hspace{5mm} i=k+1,\ldots, m \,.
\label{eq:labeling_scheme_for_propagators_and_ISPs}
\end{align}
We remark that eq.~(\ref{eq:def_generic_Feynman_integral}) does not give
a unique representation, as $D_{k+1}^{-\alpha_{k+1}} \cdots D_m^{-\alpha_m}$ can be absorbed into $N$
to form a polynomial numerator, $I(N; \alpha; D) = I\Big( N \prod_{j=k+1}^m D_j^{-\alpha_j}; \hspace{0.5mm}
(\alpha_1, \ldots, \alpha_k, \boldsymbol{0}); D\Big)$.
Nevertheless, we find it more convenient to use this notation than to fix
the rescaling invariance.

The question we wish to address is whether differential equations of the form
(\ref{eq:diff_eqs_schematic}) can be set up without
introducing integrals with squared propagators.
To this end it is convenient to make use of the Baikov representation
\cite{Baikov:1996rk} in which the integration variables are the inverse propagators
and irreducible numerator insertions, $z_\alpha = D_\alpha$ with $1 \leq \alpha \leq m$.
The associated Jacobian involves the Gram determinants
$U = \det_{i,j=1,\ldots,E} (p_i \cdot p_j)$ and $F = \det_{i,j=1,\ldots,E+L} (v_i \cdot v_j)$
where $\{ v_1, \ldots, v_{E+L} \} \equiv \{ p_1, \ldots, p_E, \ell_1, \ldots, \ell_L \}$.
Using this notation, the integral
in eq.~(\ref{eq:def_generic_Feynman_integral}) has the following Baikov representation
(up to an irrelevant kinematics-independent prefactor),
\begin{align}
I(N; \alpha; D) \hspace{0.5mm}=\hspace{0.5mm} U^\frac{E-D+1}{2} \hspace{-1mm}
\int \frac{\d z_1 \cdots \d z_m}{z_1^{\alpha_1} \cdots z_m^{\alpha_m}} F(z)^\frac{D-L-E-1}{2} N(z) \,.
\label{eq:Baikov_representation}
\end{align}
To write down differential equations of the form (\ref{eq:diff_eqs_schematic}),
we let $\big(I_1, \ldots, I_M \big)$ denote a basis of integrals and
differentiate the Baikov representation (\ref{eq:Baikov_representation})
with respect to an arbitrary external invariant $\chi$, yielding,
\begin{align}
\frac{\partial}{\partial \chi} I_j (N_j;\alpha;D)
= \frac{E-D+1}{2U} \frac{\partial U}{\partial \chi} I_j (N_j;\alpha;D) +
\frac{D{-}L{-}E{-}1}{2} I_j \Big( \frac{1}{F} \frac{\partial F}{\partial \chi} N_j; \alpha; D \Big) \,.
\label{eq:diff_eqs_in_Baikov_rep_1}
\end{align}
We observe that the $\frac{1}{F}$ factor in the
second term effectively modifies the integration measure in
eq.~(\ref{eq:Baikov_representation}), shifting
the space-time dimension from $D$ to $D-2$.

However, as proved in ref.~\cite{Bosma:2017hrk}, the second
term of eq.~(\ref{eq:diff_eqs_in_Baikov_rep_1}) can always
be expressed as a linear combination of $D$-dimensional integrals.
This follows from the fact that there exist
polynomials $(a_1, \ldots, a_m, b)$ in the $z_\alpha$ and
the external kinematical invariants such that the following relation holds,
\begin{equation}
\frac{\partial F}{\partial \chi} = \sum_{i=1}^m a_i \frac{\partial F}{\partial z_i} + bF \,,
\label{eq:Baikov_poly_ideal_membership_explicit_1}
\end{equation}
referred to as \emph{fundamental ideal membership} of $F$.

Using eq.~(\ref{eq:Baikov_poly_ideal_membership_explicit_1})
and integration by parts in each $z_i$, one finds that
eq.~(\ref{eq:diff_eqs_in_Baikov_rep_1}) takes the form
\begin{align}
\frac{\partial}{\partial \chi} I_j (N_j;\alpha;D)
= \frac{E-D+1}{2U} \frac{\partial U}{\partial \chi} I_j (N_j;\alpha;D)
+ I_j (Q_j; \alpha; D) \,,
\label{eq:diff_eqs_in_Baikov_rep_2}
\end{align}
where the insertion $Q_j$ is given by,
\begin{equation}
Q_j = \sum_{i=1}^m \left[ \alpha_i \frac{a_i N_j}{z_i}
- \frac{\partial}{\partial z_i} (a_i N_j) \right] + \frac{D{-}L{-}E{-}1}{2} b N_j \,.
\label{eq:numerator_insertion}
\end{equation}
The resulting right-hand side of eq.~(\ref{eq:diff_eqs_in_Baikov_rep_2}) thus
involves only $D$-dimensional integrals. Upon applying integration-by-parts
reductions to the the right-hand sides for each $j=1,\ldots, M$, we find
differential equations of the form (\ref{eq:diff_eqs_schematic}).

\section{Differential equations without squared propagators}\label{sec:diff_eqs_without_squared_propagators}

Having set up the differential equations (\ref{eq:diff_eqs_schematic})
in Baikov representation it is now straightforward to examine
whether it is possible to avoid introducing integrals with squared propagators.
From eqs.~(\ref{eq:diff_eqs_in_Baikov_rep_2})--(\ref{eq:numerator_insertion})
we observe that terms with positive $\alpha_i$ will produce squared
propagators for a generic polynomial $a_i$.
However, provided it is possible to choose the polynomials $a_i$ such that,
\begin{equation}
a_i = z_i b_i \hspace{5mm} \mathrm{for} \hspace{5mm} i=1,\ldots,k \,,
\label{eq:cofactor_proportionality}
\end{equation}
where $b_i$ denote polynomials, the insertion (\ref{eq:numerator_insertion})
takes the following form,
\begin{align}
Q_j = & \sum_{i=1}^k \left[ (\alpha_i{-}1) b_i N_j - z_i \frac{\partial (b_i N_j)}{\partial z_i} \right]
+ \sum_{i=k+1}^m \left[ \alpha_i \frac{a_i N_j}{z_i}
-\frac{\partial (a_i N_j)}{\partial z_i} \right] + \frac{D{-}L{-}E{-}1}{2} b N_j \,.
\label{eq:numerator_of_differentiated_integrand}
\end{align}
We observe that the only occurrence of $\frac{1}{z_i}$ is within
the sum over the range $k+1 \leq i \leq m$. However, in this range $\frac{1}{z_i^{\alpha_i}}$
occurs in the integrands with non-positive $\alpha_i$,
cf.~eqs.~(\ref{eq:labeling_scheme_for_propagators_and_ISPs})~and~(\ref{eq:Baikov_representation}).
That is, the $\frac{1}{z_i}$ in eq.~(\ref{eq:numerator_of_differentiated_integrand})
can at most introduce a propagator, but never double a propagator already present.

We conclude that no integrals with squared propagators are generated
in setting up differential equations of the form (\ref{eq:diff_eqs_schematic})
provided that the following relation holds,
\begin{equation}
\frac{\partial F}{\partial \chi} = \sum_{i=1}^k b_i z_i \frac{\partial F}{\partial z_i}
+ \sum_{i=k+1}^m a_i \frac{\partial F}{\partial z_i} + bF \,.
\label{eq:enhanced_ideal_membership}
\end{equation}
Eq.~(\ref{eq:enhanced_ideal_membership}) can be rewritten as the equivalent statement
\begin{equation}
\frac{\partial F}{\partial \chi} \in \left\langle z_1 \frac{\partial F}{\partial z_1}, \ldots,
z_k \frac{\partial F}{\partial z_k}, \frac{\partial F}{\partial z_{k+1}}, \ldots,
\frac{\partial F}{\partial z_m},F \right\rangle \,,
\label{eq:enhanced_ideal_membership_2}
\end{equation}
which we refer to as \emph{enhanced ideal membership} of $F$.

Ideal membership (\ref{eq:enhanced_ideal_membership_2}) can be determined by computing
a Gr{\"o}bner basis $\mathcal{G}$ of the ideal
on the right-hand side and then computing the remainder $r$ of $\frac{\partial F}{\partial \chi}$
after polynomial division with respect to $\mathcal{G}$.
Namely, eq.~(\ref{eq:enhanced_ideal_membership_2}) holds if and only if $r=0$.
Alternatively, one can solve explicitly for the cofactors
$(b_1, \ldots, b_k, a_{k+1}, \ldots, a_m, b)$
by starting with Ans{\"a}tze which are linear in $(z_1, \ldots, z_m)$
and iteratively allowing for cofactors of higher degree. This is
an efficient approach in practice, as cofactors are
typically of low degrees and thus lead to linear systems of manageable sizes.

\begin{figure}[!h]
\begin{center}
\includegraphics[angle=0, width=0.7\textwidth]{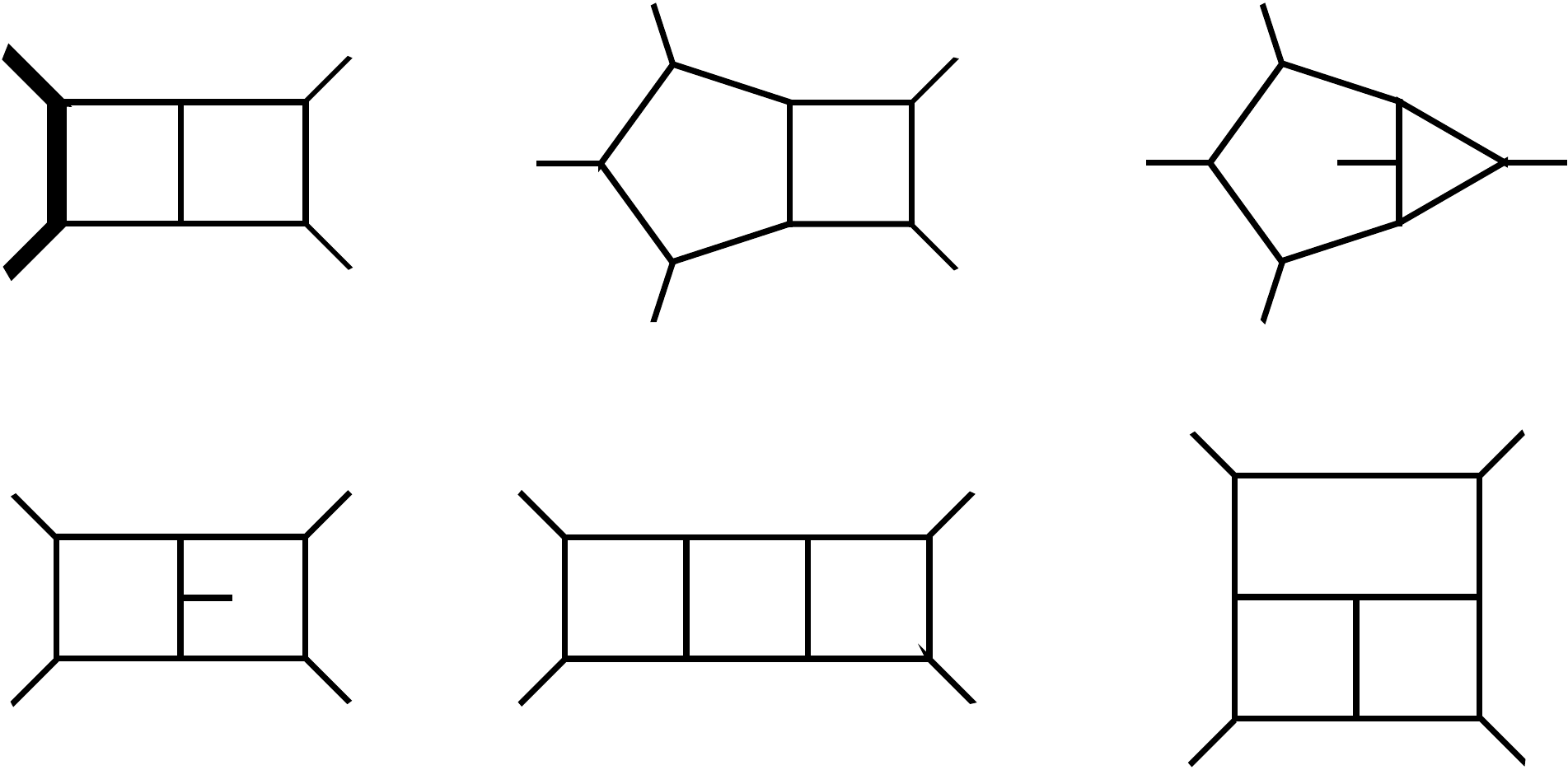}
{\vskip -0mm}
\caption{A selection of diagrams for which the enhanced
ideal membership in eq.~(\ref{eq:enhanced_ideal_membership}) holds.
The bold lines represent massive momenta and propagators.}
\label{fig:check_of_enhanced_membership}
\end{center}
\end{figure}

\begin{figure}[!h]
\begin{center}
\includegraphics[angle=0, width=0.30\textwidth]{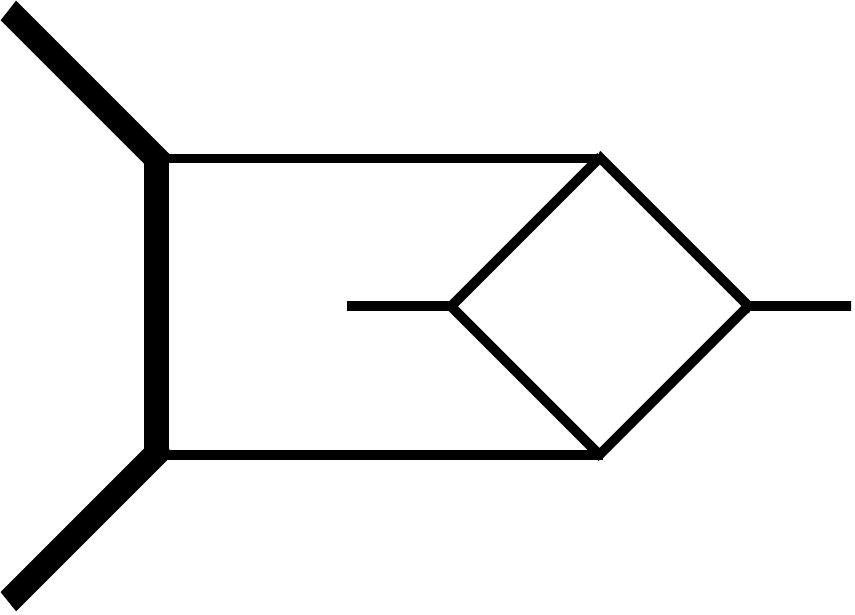}
{\vskip -0mm}
\caption{Non-planar double-box diagram. The bold lines represent massive
momenta and propagators. For this diagram, the enhanced ideal membership
(\ref{eq:enhanced_ideal_membership}) does not hold.}
\label{fig:counterexample}
\end{center}
\end{figure}

The enhanced ideal membership turns out to hold for a large class
of multi-loop integrals. Some examples are illustrated in figure~\ref{fig:check_of_enhanced_membership}.
At the same time we note that the enhanced ideal membership
(\ref{eq:enhanced_ideal_membership_2}) is not
a general property of the Baikov polynomial $F$: e.g.,
the diagram in figure~\ref{fig:counterexample} provides a counterexample.

\section{Example}\label{sec:example}

\begin{figure}[!h]
\begin{center}
\includegraphics[angle=0, width=0.35\textwidth]{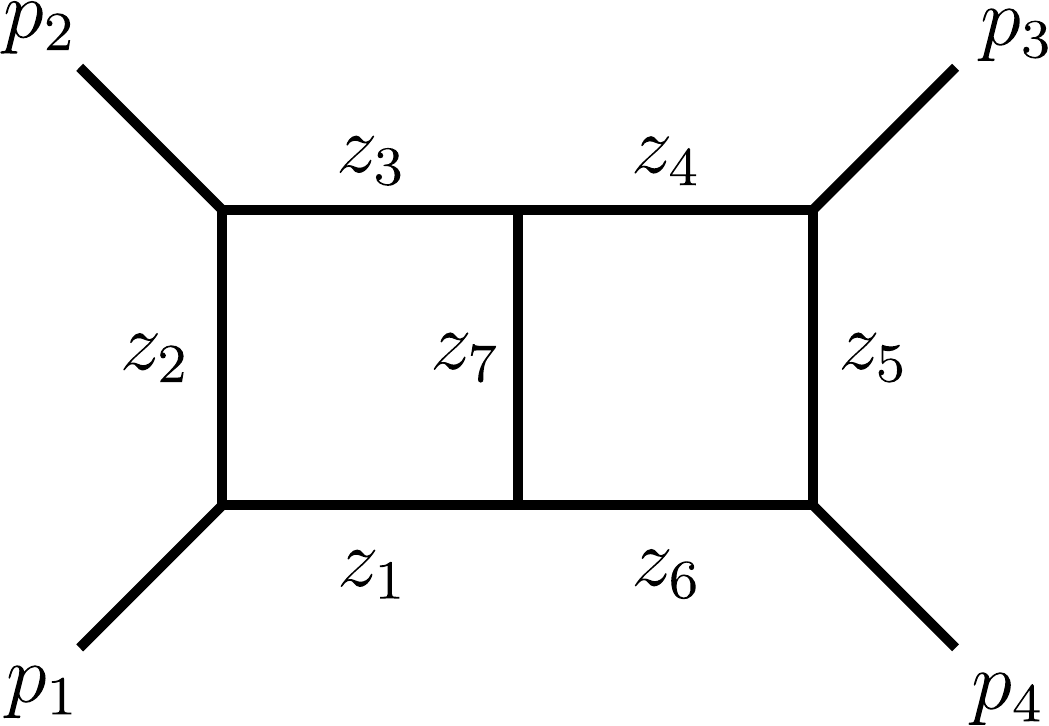}
{\vskip -0mm}
\caption{The fully massless planar double-box diagram.
All external momenta are taken to be outgoing.}
\label{fig:massless_planar_DB_z_variables}
\end{center}
\end{figure}

As an application of the formalism in %
sections~\ref{sec:Baikov_representation} and \ref{sec:diff_eqs_without_squared_propagators},
let us work out the cofactors and the differential equations
of the fully massless planar double-box diagram
shown in figure~\ref{fig:massless_planar_DB_z_variables}.

In the case at hand we have $m = 2 \cdot 3 + 2 \cdot 3 /2 = 9$
Baikov variables which we define as follows,
\begin{equation}
\begin{alignedat}{3}
z_1&=\ell_1^2\,,           \hspace{6mm}  && z_2 = (\ell_1 - p_1)^2\,,  \hspace{6mm} && z_3 = (\ell_1 - p_1 - p_2)^2\,, \\
z_4&=(\ell_2 + p_1 + p_2)^2\,,  \hspace{6mm}  && z_5 = (\ell_2 - p_4)^2\,,  \hspace{6mm} && z_6 = \ell_2^2\,, \\
z_7&=(\ell_1+\ell_2)^2\,,  \hspace{6mm}  && z_8 = (\ell_1 + p_4)^2\,,  \hspace{6mm} && z_9 = (\ell_2 + p_1)^2\,.
\end{alignedat}
\end{equation}
We wish to set up differential equations for a basis
of the vector space spanned by the diagram in figure~\ref{fig:massless_planar_DB_z_variables}
and its subdiagrams. An integral basis $\boldsymbol{\mathcal{I}}$ can be obtained
with \textsc{Azurite} \cite{Georgoudis:2016wff}. Setting $I_\alpha \equiv I(1; \alpha; D)$,
it finds
\begin{alignat}{5}
\boldsymbol{\mathcal{I}} = \Big( s^{-1 + 2\epsilon} I_{(0,1,0,0,1,0,1,0,0)},& \hspace{2mm}
&& s^{-1 + 2\epsilon} I_{(1,0,0,1,0,0,1,0,0)}, && \hspace{2mm} s^{2\epsilon} I_{(1,0,1,0,1,0,1,0,0)},&& \hspace{2mm} s^{2\epsilon} I_{(1,0,1,1,0,1,0,0,0)}, \nonumber \\
s^{1 + 2\epsilon} I_{(1,1,0,1,1,0,1,0,0)},& \hspace{2mm}
&& s^{1 + 2\epsilon} I_{(1,1,1,0,1,0,1,0,0)}, && \hspace{2mm} s^{3 + 2\epsilon} I_{(1,1,1,1,1,1,1,0,0)},&& \hspace{2mm}
s^{2 + 2\epsilon} I_{(1,1,1,1,1,1,1,-1,0)} \Big) \,.
\label{eq:planar_DB_integral_basis}
\end{alignat}
We rescaled the basis integrals $I_\alpha$ by
$s^{|\alpha| - 4 + 2\epsilon}$ to render them dimensionless.
Thus, the integrals in eq.~(\ref{eq:planar_DB_integral_basis})
depend on kinematics only through the dimensionless ratio
$\chi \equiv t/s$.

We are therefore interested in differential equations for
the basis integrals in eq.~(\ref{eq:planar_DB_integral_basis})
taken with respect to $\chi$. In the case at hand, we find
that the following property, slightly stronger than
eq.~(\ref{eq:enhanced_ideal_membership}), holds,
\begin{equation}
\frac{\partial F}{\partial \chi} =
\sum_{i=1}^9 b_i z_i \frac{\partial F}{\partial z_i} + bF \,.
\label{eq:decomposed_derivative_of_Baikov_pol}
\end{equation}
By writing Ans{\"a}tze for $(b_i,b)$ which are linear in $(z_1, \ldots, z_9)$
and solving the resulting linear system, one finds the following cofactors,
setting $\mathbf{b} = (b_1, \ldots, b_9)$,
\begin{align}
\mathbf{b} &= \left({\textstyle\frac{z_3-z_8}{\chi (\chi +1) s}},
   {\textstyle\frac{z_3 - z_8 - \chi s -s}{\chi (\chi + 1) s}},
   {\textstyle\frac{z_3 - z_8 - s}{\chi (\chi+1) s}},
   {\textstyle\frac{z_4 - z_5 - s}{\chi (\chi +1) s}},
   {\textstyle\frac{z_4 - z_5 - s}{\chi (\chi +1) s}},
   {\textstyle\frac{z_4-z_5}{\chi (\chi+1) s}},
   {\textstyle\frac{z_3 + z_4 - z_5 - z_8 - s}{\chi (\chi +1) s}},
   {\textstyle\frac{z_3 - z_8 - s}{\chi (\chi +1) s}},
   {\textstyle\frac{z_4 - z_5 - \chi s - s}{\chi (\chi +1) s}} \right) \nonumber \\[2mm]
b &= -{\textstyle\frac{2z_3 + 2z_4 - 2z_5 - 2z_8 - 2\chi s - 3s}{\chi (\chi +1) s}} \,.
\label{eq:massless_planar_four-point_cofactors}
\end{align}
We insert these expressions into eqs.~(\ref{eq:cofactor_proportionality})~and~(\ref{eq:numerator_of_differentiated_integrand})
and subsequently apply IBP reductions to the resulting right-hand sides
to find a system of differential equations of the desired form,
\begin{equation}
\frac{\partial}{\partial \chi} \boldsymbol{\mathcal{I}} (\chi, \epsilon)
= A(\chi, \epsilon) \boldsymbol{\mathcal{I}} (\chi, \epsilon) \,.
\label{eq:diff_eqs_single_ratio}
\end{equation}
The resulting coefficient matrix $A(\chi, \epsilon)$ is not particularly
illuminating. Rather than presenting its explicit form, we take
one further step \cite{Henn:2013pwa} and rotate to a new basis $\boldsymbol{\mathcal{J}}(\chi, \epsilon)$,
\begin{equation}
\boldsymbol{\mathcal{J}}(\chi, \epsilon) = U(\chi, \epsilon) \boldsymbol{\mathcal{I}}(\chi, \epsilon) \,,
\label{eq:change-of-basis_matrix}
\end{equation}
in which the coefficient matrix becomes proportional to $\epsilon$,
whereby the system is in \emph{canonical form}.

We can find a change-of-basis matrix $U$ with the desired property
by using \texttt{Fuchsia} \cite{Gituliar:2017vzm}.
Providing the coefficient matrix $A(\chi, \epsilon)$ computed
in eq.~(\ref{eq:diff_eqs_single_ratio}) as input,
it finds
\begin{align}
U &= \mathrm{diag} \Big( \textstyle{\frac{(1 - 2 \epsilon) (1 - 3 \epsilon) (2 - 3 \epsilon)}{120 \epsilon^3 \chi}},
\textstyle{\frac{(1 - 2 \epsilon) (1 - 3 \epsilon) (-2 + 3 \epsilon)}{120 \epsilon^3}},
\textstyle{\frac{(1 - 2 \epsilon) (1 - 3 \epsilon)}{24 \epsilon^2}},
\textstyle{\frac{(1 - 2 \epsilon)^2}{18 \epsilon^2}},
-\textstyle{\frac{\chi+1}{2}},
\textstyle{\frac{-1 + 2 \epsilon}{6 \epsilon}},
-\textstyle{\frac{\chi}{2}},
\textstyle{\frac{1}{2}} \Big) \,.
\end{align}
In the new basis $\boldsymbol{\mathcal{J}}(\chi, \epsilon)$,
we have the explicit differential equations
\begin{equation}
\frac{\partial}{\partial \chi} \boldsymbol{\mathcal{J}} (\chi, \epsilon)
= \epsilon \left( \frac{a_0}{\chi} + \frac{a_{-1}}{\chi + 1} \right) \boldsymbol{\mathcal{J}} (\chi, \epsilon) \,,
\end{equation}
where $a_0$ and $a_{-1}$ are matrices with integer entries,
\begin{equation}
a_0 = \begin{pmatrix}
 -2 \hspace{-0.5mm} & \hspace{-0.5mm} 0 \hspace{-0.5mm} & \hspace{-0.5mm} 0 \hspace{-0.5mm} & \hspace{-0.5mm} 0 \hspace{-0.5mm} & \hspace{-0.5mm} 0 \hspace{-0.5mm} & \hspace{-0.5mm} 0 \hspace{-0.5mm} & \hspace{-0.5mm} 0 \hspace{-0.5mm} & \hspace{-0.5mm} 0 \\
 0 \hspace{-0.5mm} & \hspace{-0.5mm} 0 \hspace{-0.5mm} & \hspace{-0.5mm} 0 \hspace{-0.5mm} & \hspace{-0.5mm} 0 \hspace{-0.5mm} & \hspace{-0.5mm} 0 \hspace{-0.5mm} & \hspace{-0.5mm} 0 \hspace{-0.5mm} & \hspace{-0.5mm} 0 \hspace{-0.5mm} & \hspace{-0.5mm} 0 \\
 0 \hspace{-0.5mm} & \hspace{-0.5mm} 0 \hspace{-0.5mm} & \hspace{-0.5mm} 0 \hspace{-0.5mm} & \hspace{-0.5mm} 0 \hspace{-0.5mm} & \hspace{-0.5mm} 0 \hspace{-0.5mm} & \hspace{-0.5mm} 0 \hspace{-0.5mm} & \hspace{-0.5mm} 0 \hspace{-0.5mm} & \hspace{-0.5mm} 0 \\
 0 \hspace{-0.5mm} & \hspace{-0.5mm} 0 \hspace{-0.5mm} & \hspace{-0.5mm} 0 \hspace{-0.5mm} & \hspace{-0.5mm} 0 \hspace{-0.5mm} & \hspace{-0.5mm} 0 \hspace{-0.5mm} & \hspace{-0.5mm} 0 \hspace{-0.5mm} & \hspace{-0.5mm} 0 \hspace{-0.5mm} & \hspace{-0.5mm} 0 \\
 -60 \hspace{-0.5mm} & \hspace{-0.5mm} -60 \hspace{-0.5mm} & \hspace{-0.5mm} 0 \hspace{-0.5mm} & \hspace{-0.5mm} 0 \hspace{-0.5mm} & \hspace{-0.5mm} -2 \hspace{-0.5mm} & \hspace{-0.5mm} 0 \hspace{-0.5mm} & \hspace{-0.5mm} 0 \hspace{-0.5mm} & \hspace{-0.5mm} 0 \\
 20 \hspace{-0.5mm} & \hspace{-0.5mm} 0 \hspace{-0.5mm} & \hspace{-0.5mm} -4 \hspace{-0.5mm} & \hspace{-0.5mm} 0 \hspace{-0.5mm} & \hspace{-0.5mm} 0 \hspace{-0.5mm} & \hspace{-0.5mm} -2 \hspace{-0.5mm} & \hspace{-0.5mm} 0 \hspace{-0.5mm} & \hspace{-0.5mm} 0 \\
 -360 \hspace{-0.5mm} & \hspace{-0.5mm} 360 \hspace{-0.5mm} & \hspace{-0.5mm} 72 \hspace{-0.5mm} & \hspace{-0.5mm} 0 \hspace{-0.5mm} & \hspace{-0.5mm} 12 \hspace{-0.5mm} & \hspace{-0.5mm} 36 \hspace{-0.5mm} & \hspace{-0.5mm} -2 \hspace{-0.5mm} & \hspace{-0.5mm} 0 \\
 540 \hspace{-0.5mm} & \hspace{-0.5mm} -360 \hspace{-0.5mm} & \hspace{-0.5mm} -90 \hspace{-0.5mm} & \hspace{-0.5mm} -9 \hspace{-0.5mm} & \hspace{-0.5mm} -18 \hspace{-0.5mm} & \hspace{-0.5mm} -36 \hspace{-0.5mm} & \hspace{-0.5mm} 1 \hspace{-0.5mm} & \hspace{-0.5mm} 1 \\
\end{pmatrix} \,,
\hspace{4mm}
a_{-1} = \begin{pmatrix}
 0 \hspace{-0.5mm} & \hspace{-0.5mm} 0 \hspace{-0.5mm} & \hspace{-0.5mm} 0 \hspace{-0.5mm} & \hspace{-0.5mm} 0 \hspace{-0.5mm} & \hspace{-0.5mm} 0 \hspace{-0.5mm} & \hspace{-0.5mm} 0 \hspace{-0.5mm} & \hspace{-0.5mm} 0 \hspace{-0.5mm} & \hspace{-0.5mm} 0 \\
 0 \hspace{-0.5mm} & \hspace{-0.5mm} 0 \hspace{-0.5mm} & \hspace{-0.5mm} 0 \hspace{-0.5mm} & \hspace{-0.5mm} 0 \hspace{-0.5mm} & \hspace{-0.5mm} 0 \hspace{-0.5mm} & \hspace{-0.5mm} 0 \hspace{-0.5mm} & \hspace{-0.5mm} 0 \hspace{-0.5mm} & \hspace{-0.5mm} 0 \\
 0 \hspace{-0.5mm} & \hspace{-0.5mm} 0 \hspace{-0.5mm} & \hspace{-0.5mm} 0 \hspace{-0.5mm} & \hspace{-0.5mm} 0 \hspace{-0.5mm} & \hspace{-0.5mm} 0 \hspace{-0.5mm} & \hspace{-0.5mm} 0 \hspace{-0.5mm} & \hspace{-0.5mm} 0 \hspace{-0.5mm} & \hspace{-0.5mm} 0 \\
 0 \hspace{-0.5mm} & \hspace{-0.5mm} 0 \hspace{-0.5mm} & \hspace{-0.5mm} 0 \hspace{-0.5mm} & \hspace{-0.5mm} 0 \hspace{-0.5mm} & \hspace{-0.5mm} 0 \hspace{-0.5mm} & \hspace{-0.5mm} 0 \hspace{-0.5mm} & \hspace{-0.5mm} 0 \hspace{-0.5mm} & \hspace{-0.5mm} 0 \\
 0 \hspace{-0.5mm} & \hspace{-0.5mm} 0 \hspace{-0.5mm} & \hspace{-0.5mm} 0 \hspace{-0.5mm} & \hspace{-0.5mm} 0 \hspace{-0.5mm} & \hspace{-0.5mm} 2 \hspace{-0.5mm} & \hspace{-0.5mm} 0 \hspace{-0.5mm} & \hspace{-0.5mm} 0 \hspace{-0.5mm} & \hspace{-0.5mm} 0 \\
 -20 \hspace{-0.5mm} & \hspace{-0.5mm} 0 \hspace{-0.5mm} & \hspace{-0.5mm} 4 \hspace{-0.5mm} & \hspace{-0.5mm} 0 \hspace{-0.5mm} & \hspace{-0.5mm} 0 \hspace{-0.5mm} & \hspace{-0.5mm} 1 \hspace{-0.5mm} & \hspace{-0.5mm} 0 \hspace{-0.5mm} & \hspace{-0.5mm} 0 \\
 360 \hspace{-0.5mm} & \hspace{-0.5mm} -720 \hspace{-0.5mm} & \hspace{-0.5mm} -36 \hspace{-0.5mm} & \hspace{-0.5mm} 18 \hspace{-0.5mm} & \hspace{-0.5mm} -12 \hspace{-0.5mm} & \hspace{-0.5mm} -36 \hspace{-0.5mm} & \hspace{-0.5mm} 2 \hspace{-0.5mm} & \hspace{-0.5mm} 2 \\
 -540 \hspace{-0.5mm} & \hspace{-0.5mm} 360 \hspace{-0.5mm} & \hspace{-0.5mm} 90 \hspace{-0.5mm} & \hspace{-0.5mm} -9 \hspace{-0.5mm} & \hspace{-0.5mm} 18 \hspace{-0.5mm} & \hspace{-0.5mm} 36 \hspace{-0.5mm} & \hspace{-0.5mm} -1 \hspace{-0.5mm} & \hspace{-0.5mm} -1 \\
\end{pmatrix} \,.
\end{equation}
Thus we have derived differential equations of the type (\ref{eq:diff_eqs_schematic})
for the basis integrals in eq.~(\ref{eq:planar_DB_integral_basis})
and achieved a canonical form of the system
without introducing integrals with squared propagators in intermediate stages.

\section{Conclusions}\label{sec:conclusions}

Differential equations of the form (\ref{eq:diff_eqs_schematic})
provide a powerful method for computing multi-loop integrals.
In practice, setting up such equations for
multi-scale integrals is non-trivial, as the step of
expressing the derivatives of the integrals in the basis
requires integration-by-parts (IBP) reductions which are
computationally intensive to generate.
Refs.~\cite{Gluza:2010ws,Ita:2015tya,Larsen:2015ped,Boehm:2018fpv}
provide a simplified approach to IBP reductions
where integrals with squared propagators are avoided
in intermediate stages, thus producing significantly
smaller linear systems to be solved.

In these proceedings, based on ref.~\cite{Bosma:2017hrk},
we have addressed the question whether it is possible
to set up differential equations of the form (\ref{eq:diff_eqs_schematic})
without introducing integrals with squared propagators in
intermediate stages, so that the formalism of
refs.~\cite{Gluza:2010ws,Ita:2015tya,Larsen:2015ped,Boehm:2018fpv}
can be applied.

We have shown that a sufficient condition is that the
Baikov polynomial $F$ satisfies eq.~(\ref{eq:enhanced_ideal_membership}).
This condition holds for a large class of multi-loop diagrams,
including highly non-trivial loop diagrams whose differential equations
are not attainable with standard methods. A sample is illustrated
in figure~\ref{fig:check_of_enhanced_membership}.
At the same time, we have identified a counterexample to
eq.~(\ref{eq:enhanced_ideal_membership}), shown in figure~\ref{fig:counterexample}.
An interesting open problem is therefore to classify the diagrams for which
the enhanced ideal membership property (\ref{eq:enhanced_ideal_membership})
holds. Another interesting problem is to find closed formulas
for the cofactors in eq.~(\ref{eq:enhanced_ideal_membership}).

\bibliographystyle{h-physrev}

\bibliography{DiffEqsWithoutSquaredProps}

\begin{thebibliography}{10}

\bibitem{Laporta:2000dc}
S.~Laporta,
\newblock Phys. Lett. {\bf B504}, 188 (2001), hep-ph/0102032.

\bibitem{Laporta:2001dd}
S.~Laporta,
\newblock Int. J. Mod. Phys. {\bf A15}, 5087 (2000), hep-ph/0102033.

\bibitem{Anastasiou:2004vj}
C.~Anastasiou and A.~Lazopoulos,
\newblock JHEP {\bf 07}, 046 (2004), hep-ph/0404258.

\bibitem{Smirnov:2008iw}
A.~V. Smirnov,
\newblock JHEP {\bf 10}, 107 (2008), 0807.3243.

\bibitem{Smirnov:2014hma}
A.~V. Smirnov,
\newblock Comput. Phys. Commun. {\bf 189}, 182 (2014), 1408.2372.

\bibitem{Studerus:2009ye}
C.~Studerus,
\newblock Comput. Phys. Commun. {\bf 181}, 1293 (2010), 0912.2546.

\bibitem{vonManteuffel:2012np}
A.~von Manteuffel and C.~Studerus,
\newblock (2012), 1201.4330.

\bibitem{Lee:2012cn}
R.~N. Lee,
\newblock (2012), 1212.2685.

\bibitem{Maierhoefer:2017hyi}
P.~Maierh{\"o}fer, J.~Usovitsch, and P.~Uwer,
\newblock Comput. Phys. Commun. {\bf 230}, 99 (2018), 1705.05610.

\bibitem{Gluza:2010ws}
J.~Gluza, K.~Kajda, and D.~A. Kosower,
\newblock Phys.Rev. {\bf D83}, 045012 (2011), 1009.0472.

\bibitem{Ita:2015tya}
H.~Ita,
\newblock Phys. Rev. {\bf D94}, 116015 (2016), 1510.05626.

\bibitem{Larsen:2015ped}
K.~J. Larsen and Y.~Zhang,
\newblock Phys. Rev. {\bf D93}, 041701 (2016), 1511.01071.

\bibitem{vonManteuffel:2014ixa}
A.~von Manteuffel and R.~M. Schabinger,
\newblock Phys. Lett. {\bf B744}, 101 (2015), 1406.4513.

\bibitem{Bern:2017gdk}
Z.~Bern, M.~Enciso, H.~Ita, and M.~Zeng,
\newblock (2017), 1709.06055.

\bibitem{Boehm:2018fpv}
J.~B{\"o}hm, A.~Georgoudis, K.~J. Larsen, H.~Sch{\"o}nemann, and Y.~Zhang,
\newblock (2018), 1805.01873.

\bibitem{Chawdhry:2018awn}
H.~A. Chawdhry, M.~A. Lim, and A.~Mitov,
\newblock (2018), 1805.09182.

\bibitem{Kotikov:1990kg}
A.~V. Kotikov,
\newblock Phys. Lett. {\bf B254}, 158 (1991).

\bibitem{Kotikov:1991pm}
A.~V. Kotikov,
\newblock Phys. Lett. {\bf B267}, 123 (1991).

\bibitem{Bern:1993kr}
Z.~Bern, L.~J. Dixon, and D.~A. Kosower,
\newblock Nucl. Phys. {\bf B412}, 751 (1994), hep-ph/9306240.

\bibitem{Remiddi:1997ny}
E.~Remiddi,
\newblock Nuovo Cim. {\bf A110}, 1435 (1997), hep-th/9711188.

\bibitem{Gehrmann:1999as}
T.~Gehrmann and E.~Remiddi,
\newblock Nucl. Phys. {\bf B580}, 485 (2000), hep-ph/9912329.

\bibitem{Henn:2013pwa}
J.~M. Henn,
\newblock Phys. Rev. Lett. {\bf 110}, 251601 (2013), 1304.1806.

\bibitem{Papadopoulos:2014lla}
C.~G. Papadopoulos,
\newblock JHEP {\bf 07}, 088 (2014), 1401.6057.

\bibitem{Ablinger:2015tua}
J.~Ablinger {\em et~al.},
\newblock Comput. Phys. Commun. {\bf 202}, 33 (2016), 1509.08324.

\bibitem{Frellesvig:2017aai}
H.~Frellesvig and C.~G. Papadopoulos,
\newblock JHEP {\bf 04}, 083 (2017), 1701.07356.

\bibitem{Zeng:2017ipr}
M.~Zeng,
\newblock JHEP {\bf 06}, 121 (2017), 1702.02355.

\bibitem{Bosma:2017hrk}
J.~Bosma, K.~J. Larsen, and Y.~Zhang,
\newblock Phys. Rev. {\bf D97}, 105014 (2018), 1712.03760.

\bibitem{Baikov:1996rk}
P.~A. Baikov,
\newblock Phys. Lett. {\bf B385}, 404 (1996), hep-ph/9603267.

\bibitem{Georgoudis:2016wff}
A.~Georgoudis, K.~J. Larsen, and Y.~Zhang,
\newblock Comput. Phys. Commun. {\bf 221}, 203 (2017), 1612.04252.

\bibitem{Gituliar:2017vzm}
O.~Gituliar and V.~Magerya,
\newblock Comput. Phys. Commun. {\bf 219}, 329 (2017), 1701.04269.

\end{thebibliography}


\end{document}